\documentclass[twocolumn]{revtex4-1}

\usepackage[innercaption]{sidecap}
\usepackage{amssymb}
\usepackage{graphicx}
\usepackage{graphicx,epstopdf}
\usepackage{xcolor, soul}
\usepackage{dcolumn}
\usepackage{bm}
\usepackage{mathrsfs} 
\usepackage{amsmath} 
\usepackage[colorlinks=true,linkcolor=blue,citecolor=blue]{hyperref}%
\usepackage{fontenc}
\usepackage{float}
\usepackage{amsthm}
\usepackage{subfigure}
\usepackage{color}
\usepackage{ragged2e}
\usepackage{verbatim}
\usepackage{enumerate}
\usepackage{hyperref}
\def\be{\begin{equation}}
\def\ee{\end{equation}}
\def\bea{\begin{eqnarray}}
\def\eea{\end{eqnarray}}
\def\bfg{\begin{figure}[H]}
\def\efg{\end{figure}}

\begin{document}

\title{Emergence of first-order and second-order phase transitions in a cyclic ecosystem exposed to environmental impact}

\author{Sirshendu Bhattacharyya} 

\email{sirs.bh@gmail.com}

\address{Department of Physics, Raja Rammohun Roy Mahavidyalaya, Radhanagar, Hooghly 712406, India}


\begin{abstract}
\noindent
A cyclically dominating three-species ecosystem, modeled within the framework of rock-paper-scissor game, is studied in presence of natural death and an effect of the environment. The environmental impact is parameterized along with the death rates in the rate equation of the species densities. Monte-Carlo simulations on this system reveal that the population density bears the signature of first-order and second-order phase transitions in different regimes of the parameters representing natural deaths of the species. The connection of this phenomena with the phase transitions is also supported by the behavior of the basin entropy calculated for the system. The density of total population evidently becomes an order parameter with respect to the change in environmental impact on the system.\\

\end{abstract}

\maketitle

\section{Introduction}
\label{intro}
\noindent
A balanced ecosystem consists of a number of species existing simultaneously. However, within such an ecosystem each of the constituent species has to fight against numerous adversities to avoid extinction {\citep{hallmann2014,sutherland2015,buckley2015}.} The adversities that may put the existence of the species at risk, can have different mode of operations. Sometimes one or a few species within the system become vulnerable towards such effect. For example, outbreak of an infectious disease suddenly increases the death rate of some species. On the other hand the adversities can have global impact on the entire ecosystem. Any kind of harmful environmental effect can be thought of as such an agent. Climate change, global warming, environmental pollution are some of them that imparts direct and indirect effects on the species {\citep{hallmann2014,mooney2009,chaudhary2021}.} For example, the risk of species extinction increases with the increment of warming of the environment. Increasing pollution also affects the lifespan of each species in the ecosystem. The role of these adverse parameters in the maintenance of biodiversity is evidently an important theme of study.

\par The game theoretic approach towards the issue of preserving biodiversity opens an avenue for mathematical study in this regard \citep{may1972,hauert2005,szabo2007,freyPhysica2010}. {To study an ecosystem accommodating diverse populations, cyclically interacting models particularly seem to be quite useful as they demonstrate coexistence effectively} {\citep{lotka1920,volterra1926,itoh1973,may1975,szolnoki2014,dobramysl2018,szolnoki2020,zhong2022,serrao2017,menezes2019}.}   { Mapping such model onto a lattice further opens up the possibility of carrying out stochastic simulation \citep{tainaka1988}.} The simplest form of this model is the rock-paper-scissor (RPS) model which contains three species having cyclic inter-species predation and prey \citep{kerr2002,reichenbach2007mobility}. The RPS model has been studied, both analytically and numerically, from different perspectives to understand various effects acting on and emerging from an ecosystem. {In general the state of coexistence are maintained by a set of parameters. These parameters control the increase or decrease of the population \citep{siepielski2010}.} Combined effect of these parameters according to their strengths gives rise to coexistence. {Various investigations have been done within and beyond RPS framework to understand coexistence from different outlooks like the impact of mobility \citep{reichenbach2007mobility}, spatial patterns formed during coexistence \citep{avelino2012,avelino2018,reichenbachJTB2008,pal2021}, the effect of mutation \citep{mobilia2010,kleshnina2022,mittal2020} etc.} {From a mathematical point of view, the models can be described by non-linear coupled rate equations which are mean-field approximations to the stochastic system. The interactions like predation, reproduction, different types of death etc. act as rate constants.} The exact solution of these rate equations are unfeasible most of the time unless the system is excessively simple which is again unrealistic. However the stability of the system with respect to the rate constants or the properties of the population fluctuations can be studied by {statistical} mechanical treatment \citep{goel1971}.

\par Out of different parameters (implemented as rate constants in the rate {equations}) the rate of death of a species has been found to play an important role in preserving coexistence \citep{holt1985,abrams2001,mittelbach2004,gross2009,metz2010}.
The impacts of conditional \citep{avelino2019} and natural deaths \citep{bhattacharyya2020,islam2022} have been studied in spatial and non-spatial models. Even in presence of other inter-species and intra-species interactions like predation and reproduction, the natural death rate has been shown to be the majorly {contributing} factor in bringing an ecosystem to coexistence or single-species existence \citep{bhattacharyya2020}. In addition to the internal effects, an ecosystem faces external impacts as well.

\par Another important aspect of this system which has been recently revealed through interdisciplinary study is its connection with the topological phases of matter \citep{tainaka1991,knebel2020,yoshida2021,yoshida2022}. {In the field of statistical and condensed matter physics, the topological phase transition has a pivotal role in understanding the phases of matters that lack in local symmetry and cannot be explained by Landau theory.} { Surprisingly, rock-paper-scissor model sometimes carry signatures of topological phase transitions under certain influences. One of the earliest studies on this matter finds such a transition by defining vortices as the points where three species meet \citep{tainaka1991}. In another study, the RPS models driven by certain parameters show behavioral changes which are characteristically similar to topological phase transitions \citep{knebel2020}.}

\par {Apart from the topological phase transition, ecosystems mimic the phase transitions classified by Ehrenfest \citep{jaeger1998} as well.} These types of phase transitions are explainable in the Landau framework. 
The general concept of these phase transitions is manifested in diverse fields including biological systems \citep{yashroy1990,dmitri2000}, biological networks \citep{krotov2014} or the formation of flocks and patterns \citep{bialek2014} etc. Talking about phase transition in particular, the idea that biological systems may possess a criticality is quite well-established \citep{bak2013}. Experimental evidences are there in the groups of human, mice and trees that correlations increases under stress when the group leave the comfort zone. After the stress crosses some critical value, the correlations decrease and the group goes to crisis zone \citep{gorban2010}. This scenario also depicts the phenomena of 2nd order phase transition. { Studies on active to absorbing phase transitions have also been made on Lotka-Volterra systems \citep{satulovsky1994}. {Another interesting work reports phase transition originated through a transcritical bifurcation in a metapopulation model under mean-field diffusive coupling \citep{banerjee2015}.} Phase transition is therefore a widely investigated phenomena in the field of population dynamics \citep{tauber2014}.} {Not only the equilibrium phase transitions, studies have been made to obtain dynamical phase diagram also for RPS systems out of equilibrium in presence of mobility \citep{venkat2010}.}

\par Motivated by these interdisciplinary studies, our present work searches for further evidences that may connect the system's behavior with stastical and condensed matter phenomena. An ecosystem with cyclically predating species along with reproduction and natural death rates is exposed to a global impact to which all the species are equally vulnerable. The system is driven to coexistence or single species survival or even null population for different values of interaction parameters. As a result of changing the strength of this impact, the population of the system shows discontinuous and continuous phase transitions depending on the internal setup of inter-species interactions. {However it may be noted that change of the impact does not mean a varying environment. Rather, we consider a constant environment (in space and time) but with multiple different impacts on the species.}  { We identify an order-parameter-like behavior in the density of total population in these two types of phase transitions.} 
{It is always interesting to observe the phenomena of phase transitions beyond the field of statistical and condensed matter physics. Our work demonstrates such phenomena in a simple $3$-species RPS model and also draws a phase diagram having tricritical point.}

\par The remaining part of the paper is divided in three more sections. Section~\ref{model} describes the model and the method of numerical simulation. The related rate equations and the corresponding stabilities have also been discussed here. Section~\ref{result} demonstrates the results which are categorized as that of the discontinuous and the continuous phase transitions. Finally in section~\ref{conclu} we conclude with the discussion and the viewpoint on the obtained results.

\section{The model and the rate equations}
\label{model}
\noindent
We construct a 2-dimensional lattice {consisting} of three species interacting in a cyclic pattern through nearest neighbour interaction. It mimics the RPS model for evolutionary dynamics. In addition, the elements of the species may increase their numbers by reproduction. The death parameter tends to decrease the population on the other hand. Here the act of death (due to aging or due to accident) is not actually an interaction in the sense that it does not depend on the presence or absence  of any other member of the system. As the death rate creates a vacant site, we follow the May-Leonard formulation \citep{may1975} where vacant sites are also there in addition to three species and the total number of individuals is not conserved. The predation strategy creates a vacant site in the adjacent neighbor whereas the reproduction fills a vacant site with an individual \citep{avelinoPRE2019}. Let us assume that the densities of species $A$, $B$ and $C$ are $\rho_a$, $\rho_b$ and $\rho_c$ respectively. These three densities, along with the density of vacant sites ($\rho_v$), follow the conservation rule: $\rho_a + \rho_b + \rho_c + \rho_v = 1$. 
We can write the cyclic predation strategy in the following way
\be 
\lbrace A, B, C \rbrace + \lbrace B, C, A \rbrace \xrightarrow{\text{rate~~} {p_{\lbrace a, b, c \rbrace}}} \lbrace A, B, C \rbrace + V
\label{p}
\ee
\noindent
where V denotes a vacant site. The reproduction equations may be written as
\be 
\lbrace A, B, C \rbrace + V \xrightarrow{\text{rate~~} {r_{\lbrace a, b, c \rbrace}}} 2\;\lbrace A, B, C \rbrace
\label{r}
\ee 
{In addition to these two interactions, we take into account the effect of the environment on the existence of the species. In our model we assume that the external impact due to the environment is homogeneous and acts on all the three constituent species by increasing or decreasing their death rates. This may be represented as}
\be 
\lbrace A, B, C \rbrace \xrightarrow{\text{rate~~} {\varepsilon \cdot d_{\lbrace a, b, c \rbrace}}} V
\label{ed}
\ee
where $d_{a,b,c}$ are the death rates of the three species and the parameter $\varepsilon$ influences all the death rates in similar manner.

\par We plan to study the stochastic dynamics primarily by {Monte-Carlo} simulation on the above model in a 2-dimensional lattice with periodic boundary conditions applied. Our previous works \citep{bhattacharyya2020,islam2022} report the significance of the death rates in the maintenance of coexistence and emergence of the spiral patterns in spatial distributions. As discussed earlier, an individual has to struggle with the environment in addition to confronting other species. We focus on those effects of environment that endanger the existence of life in an ecosystem. Therefore, in this present work we mimic this effect by the parameter $\varepsilon$, which controls the natural death of all the constituent species and investigate the dynamics of the population.

\par
\begin{figure*}[t!]
	\begin{center}
	\includegraphics[scale=0.5]{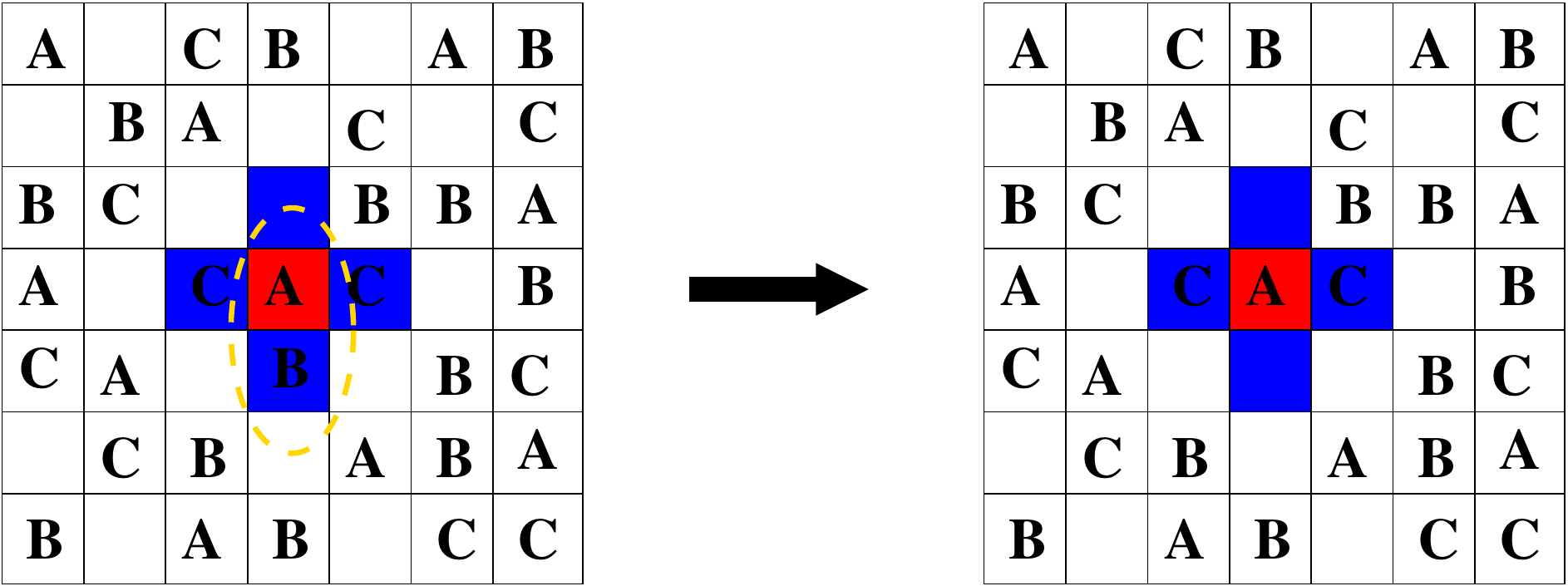}
	\end{center}
	\caption{{Schematic representation of the lattice and one arbitrary Monte-Carlo step: Species $A$, $B$, $C$ and vacant sites are distributed in the $2$-D lattice. Then one site is chosen randomly (marked in red color). The four nearest neighbors of the chosen site are marked in blue color. Out of these four, one is chosen for interaction (encircled by yellow {dashed} line). Then the interaction takes place with corresponding probability. In this figure, $A$ predates its chosen nearest neighbor $B$.}}
	\label{pred}
\end{figure*}

 The {Monte-Carlo} simulation is based on the interaction with any one of four individuals in the von Neumann neighborhood at one {Monte-Carlo} step. The simulation starts from a randomly constructed initial configuration of three species. At each Monte-Carlo step, we first select a primary non-empty site randomly and then one of its four nearest neighbors. For a non-empty nearest neighbor, predation-prey is performed with probabilities, $p_{a,b,c}$ as the case may be. On the other hand, reproductions take place with probabilities $r_{a,b,c}$ if the chosen nearest neighbor site is empty. A typical case of predation is shown schematically in {figure~\ref{pred}}. Along with the predation and reproduction, the individual residing in the primary site abolish with a probability $\varepsilon\cdot d_{a,b,c}$ making the corresponding site vacant. {Technically at each {Monte-Carlo} step, we randomly select a non-empty site as a primary site and also select one site out of
its four nearest neighbors randomly. Then one of the three actions i.e, death, predation and reproduction is performed at a step with corresponding probabilities assigned. As a result, the species in the primary site may abolish with probability $d_{a,b,c}$, or predation-prey may occur for non-empty nearest-neighbor site with probability $p_{a,b,c}$, or the individual in the primary site may attempt reproduction with probability $r_{a,b,c}$ in case the nearest neighbor is found to be empty. After each Monte-Carlo step, the chosen site and the chosen nearest neighbor are updated accordingly.} {The results obtained here are averaged over $10$ different initial configurations.} The time unit of our calculation is defined by $N$ {Monte-Carlo} steps, where $N$ is the system size { (i.e. the lattice size which is also equal to the sum of the numbers of
three species and vacancy, because each site can be occupied by one individual only or remain vacant)}.

{The temporal evolution of the densities of the three species can also be obtained from the rate equations derived after mean-field approximation.} We can construct them in the form of coupled non-linear differential equations involving predation, reproduction, death and environmental impact as rate constants. These differential equations looks like
\begin{eqnarray}
\dfrac{\partial \rho_a}{\partial t} &=& \rho_a(t) \left [ {r_{a}} \rho_v(t)- p_{c}\rho_c(t)-\varepsilon\cdot d_{a}\right] \nonumber \\
\dfrac{\partial  \rho_b}{\partial t} &=& \rho_b(t) \left [ r_{b} \rho_v(t)- p_{a} \rho_a(t)-\varepsilon\cdot d_{b}\right] \nonumber \\
\dfrac{\partial \rho_c}{\partial t} &=& \rho_c(t) \left [ r_{c} \rho_v(t)- p_{b} \rho_b(t)-\varepsilon\cdot d_{c}\right]
\label{rate-eq}
\end{eqnarray}
{where $\rho_a(t)$, $\rho_b(t)$, $\rho_c(t)$ and $\rho_v(t)$  are the densities of species $A$, $B$, $C$ and vacancy respectively at time $t$ and $\rho_v(t) = 1 - \rho_a(t) - \rho_b(t) - \rho_c(t)$.} We get three sets of fixed points for equation~(\ref{rate-eq}). The following sets of fixed points are obtained for the above rate equations \citep{bhattacharyya2020}
\begin{enumerate}
\item[I.] Trivial fixed point: $\rho_a=\rho_b=\rho_c=0$.
\item[II.] Single species fixed points:
\begin{enumerate}
\item[(a)] $\rho_a=1-\dfrac{\varepsilon\cdot d_a}{r_a}$, $\rho_b=\rho_c=0$.
\item[(b)] $\rho_b=1-\dfrac{\varepsilon\cdot d_b}{r_b}$, $\rho_a=\rho_c=0$.
\item[(c)] $\rho_c=1-\dfrac{\varepsilon\cdot d_c}{r_c}$, $\rho_a=\rho_b=0$.
\end{enumerate}
\item[III.] Coexistence fixed point:\\ $\rho_a =\dfrac{r_b}{p_a}Q - \dfrac{\varepsilon\cdot d_b}{p_a}$,\\ $\rho_b =\dfrac{r_c}{p_b}Q - \dfrac{\varepsilon\cdot d_c}{p_b}$,\\ $\rho_c =\dfrac{r_a}{p_b}Q - \dfrac{\varepsilon\cdot d_a}{p_b}$\\
where $Q=\dfrac{1+\varepsilon\cdot \left(\frac{d_a}{p_c}+\frac{d_b}{p_a}+\frac{d_c}{p_b}\right)}{1+\left(\frac{r_a}{p_c}+\frac{r_b}{p_a}+\frac{r_c}{p_b}\right)}$.
\end{enumerate}
The system moves to a feasible fixed point depending on the values of the parameters. The dynamics obtained from the {Monte-Carlo} simulation also equilibrates to the same densities here.

\section{Results}
\label{result}
\noindent
The model ecosystem evolving with assigned predation, reproduction and death rates with environmental impact shows different temporal behaviors of the species densities in different regimes of the values of parameters. We denote the density of the entire population as $\rho = \rho_a + \rho_b + \rho_c$ and observe its behavior in different situations. The reason behind such observation will be more clear subsequently when the phase transitions will be indicated by this quantity with respect to the external impact $\varepsilon$. We mainly focus in two different regimes depending on the death rates of the species, keeping the predation rates and reproduction rates same for all the three species, i.e., $p_a=p_b=p_c=0.2$ and $r_a=r_b=r_c=0.4$ throughout the entire work. {We keep the system size $N=100\times 100$ in all the results shown here. However, keeping in mind the impact of finite-size effects on dynamic phase transitions \citep{korniss2002} we checked that in our case, further  increase of system size only reduces the small fluctuations in the density profile and does not affect the key observations.}

\subsection{Transition from coexistence to single species survival state}
\label{1st-order}

The first case deals with the death rates {maintaining} the criteria: $d_a<r_a$, $d_a\leq d_b$, $d_a<d_c$. The reason behind taking such values lies in the fact that this criteria enforces the survival of single species $A$ in absence of the environmental effect \citep{bhattacharyya2020}. However, in the present {observation}, we find that even after this criteria being fulfilled, the system shows coexistence for small values of $\varepsilon$.
\par
\begin{figure*}[t!]
	\includegraphics[scale=0.3]{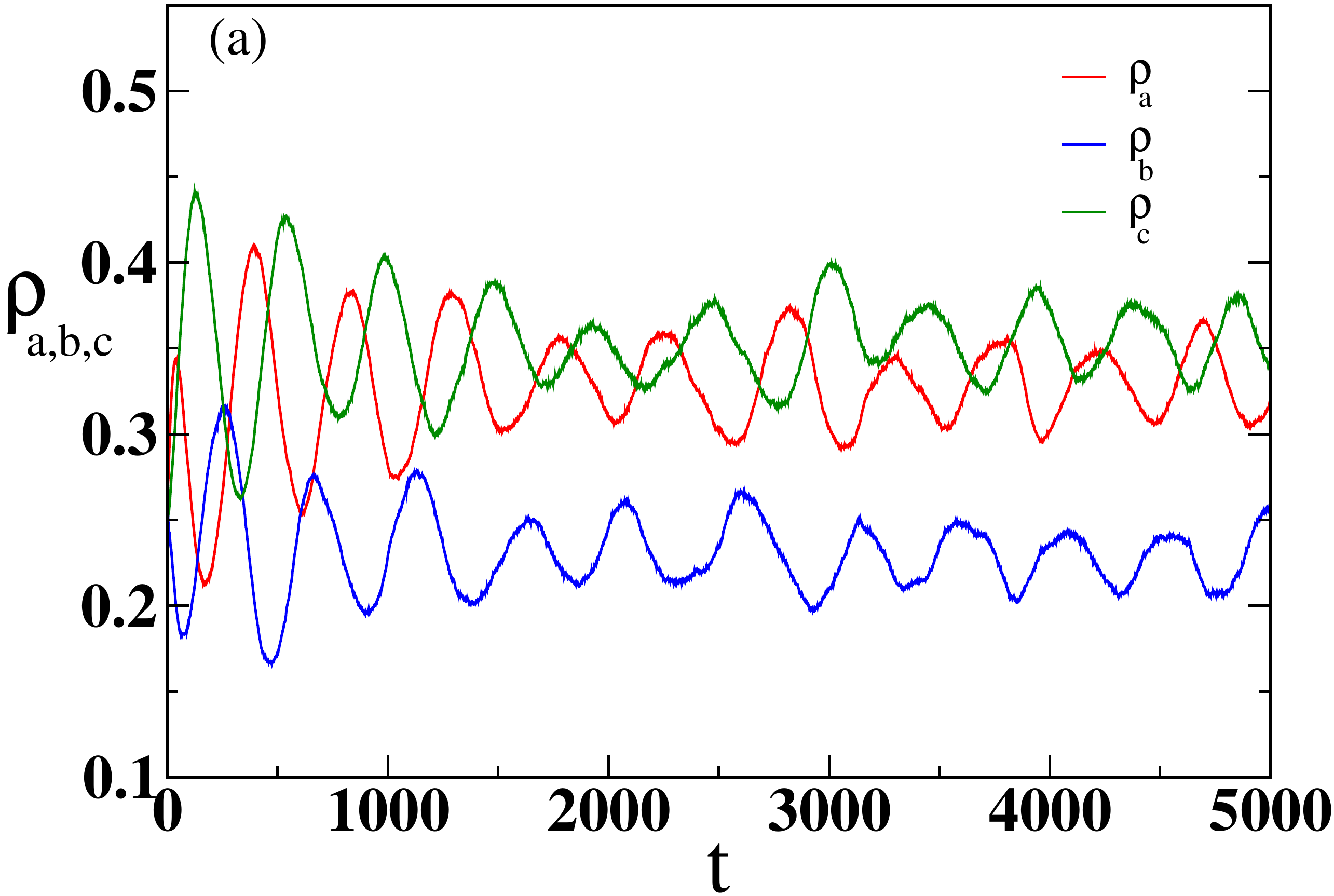}
	\includegraphics[scale=0.3]{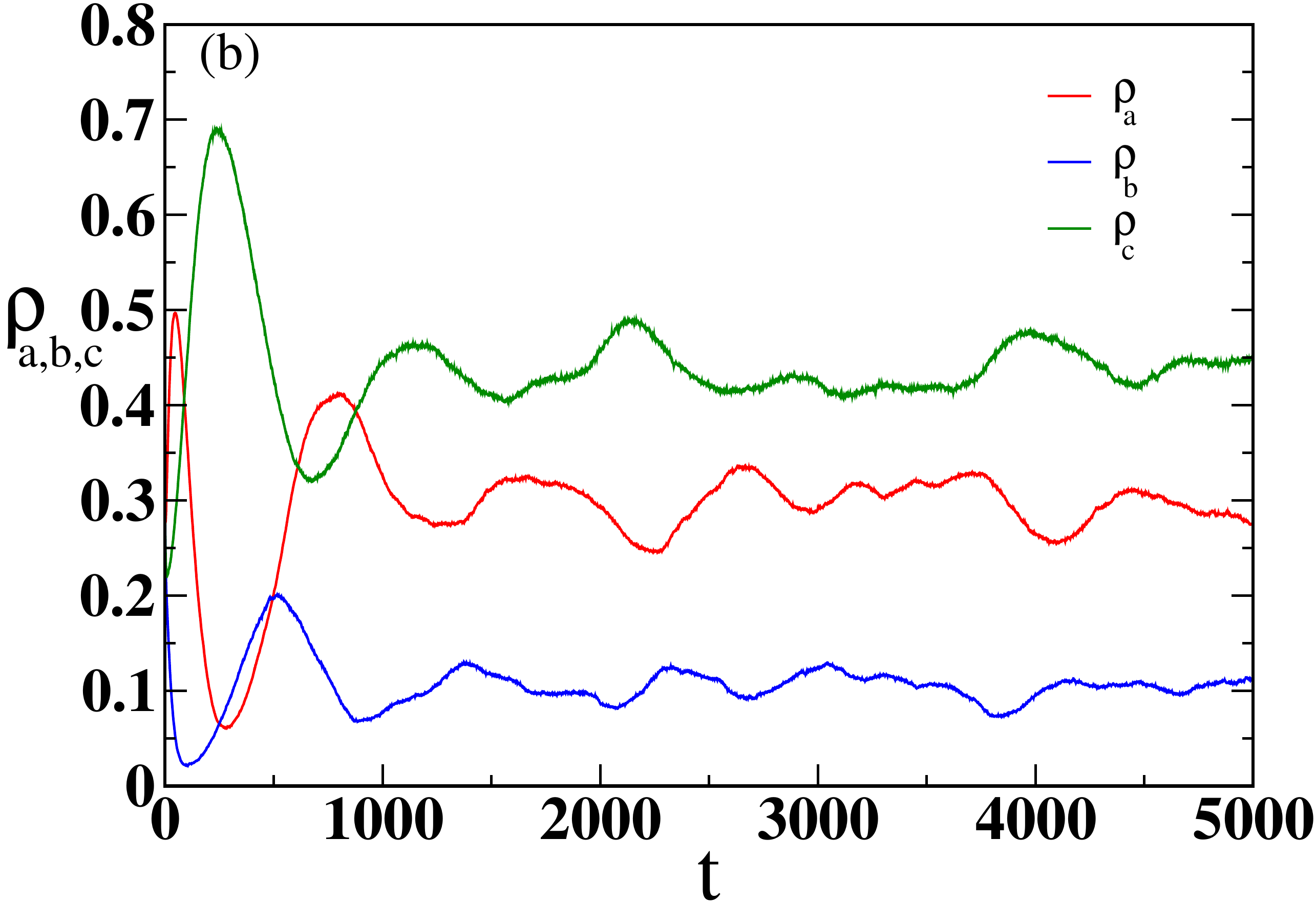}\\
	\includegraphics[scale=0.3]{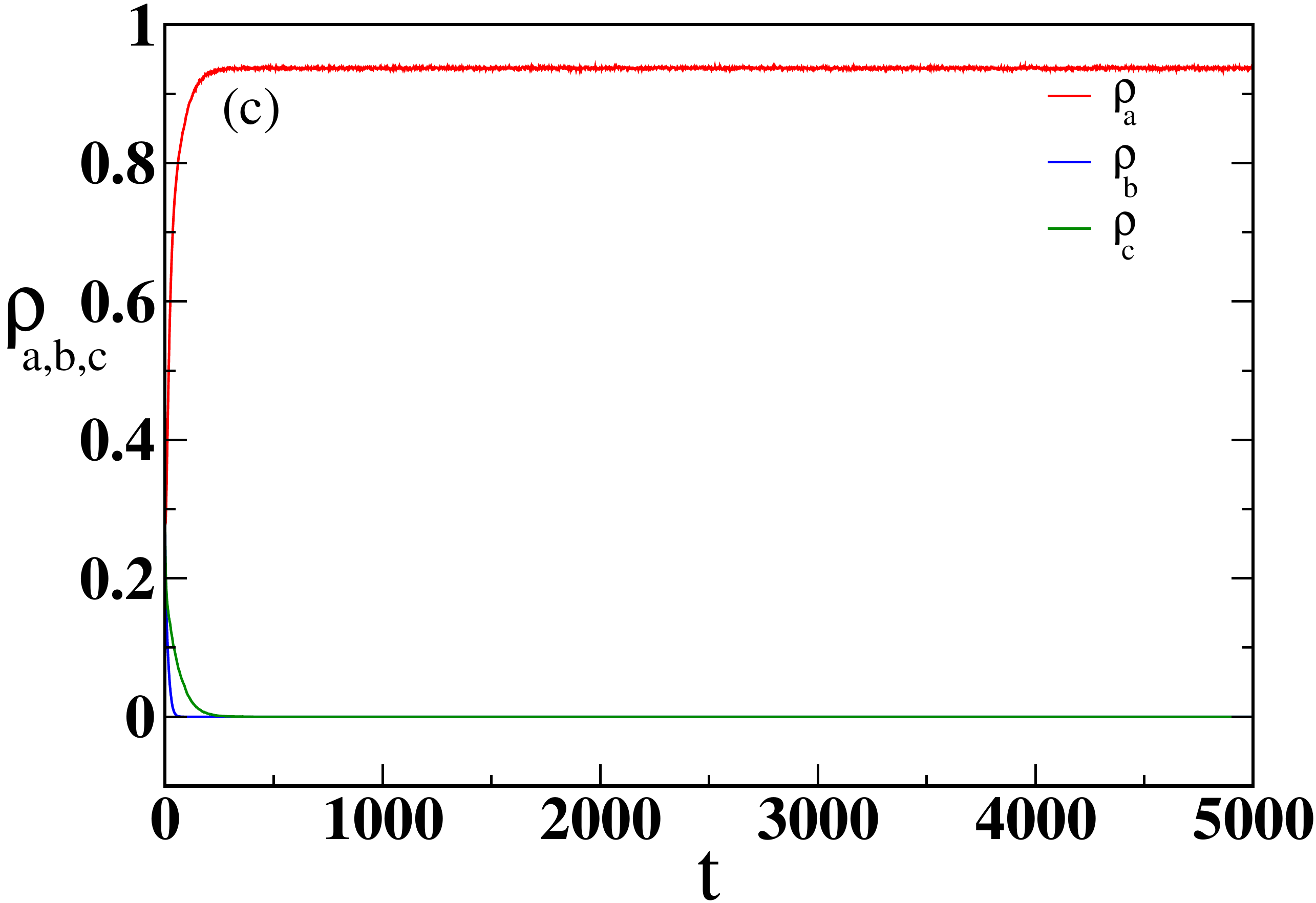}
	\includegraphics[scale=0.3]{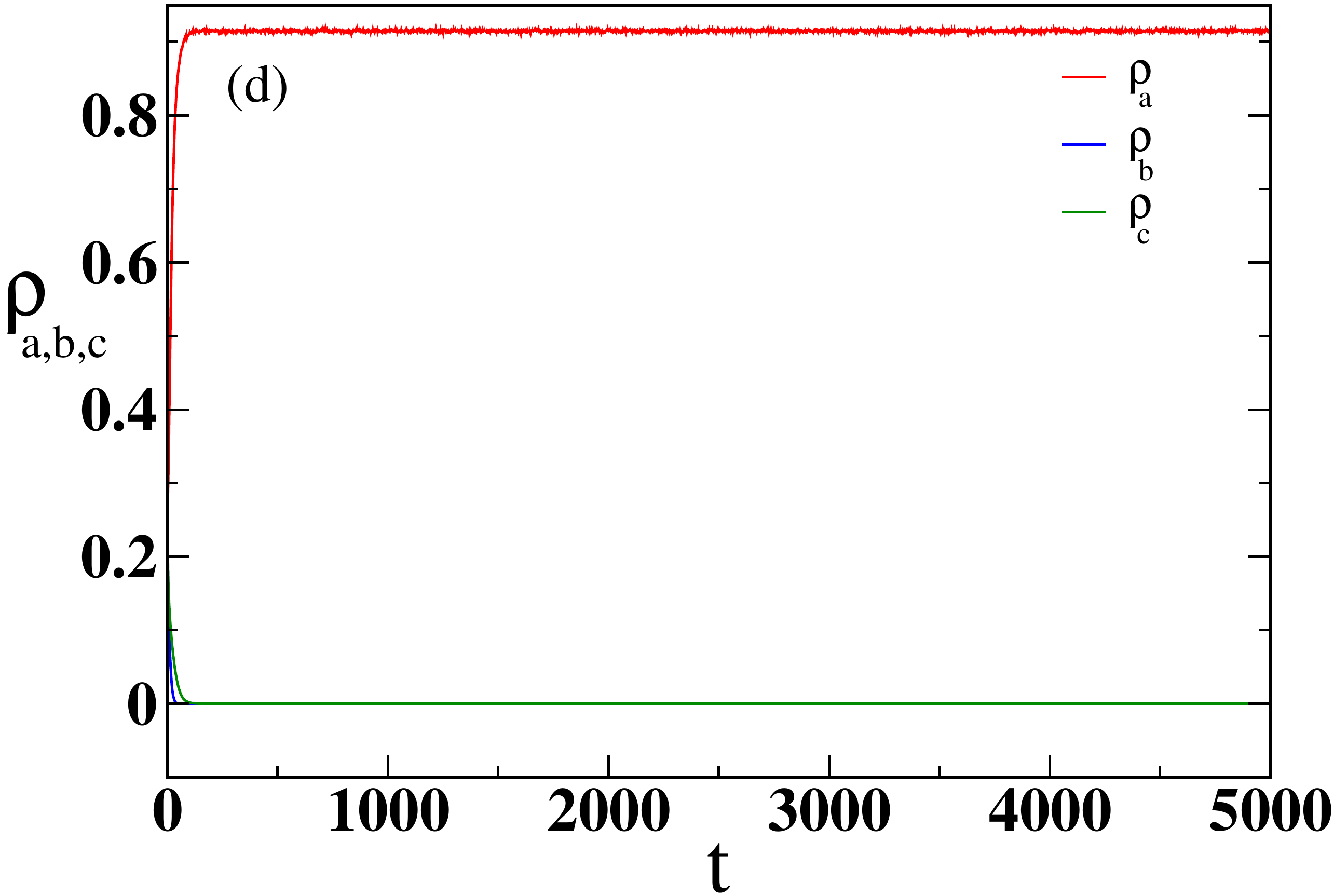}
	\caption{Density vs time plot for different $\varepsilon$ values, keeping other parameters fixed: $p_a=p_b=p_c=0.2$; $r_a=r_b=r_c=0.4$; $d_a=0.02$, $d_b=d_c=0.1$. (a)-(b) Coexistence is observed for $\varepsilon=0.2$ and $0.6${,} respectively. (c)-(d) Survival of single species $A$ is shown for $\varepsilon=1.2$ and $1.6$.}
	\label{density-single}
\end{figure*}

\begin{figure*}[t!]
	\begin{center}
	\includegraphics[scale=0.6]{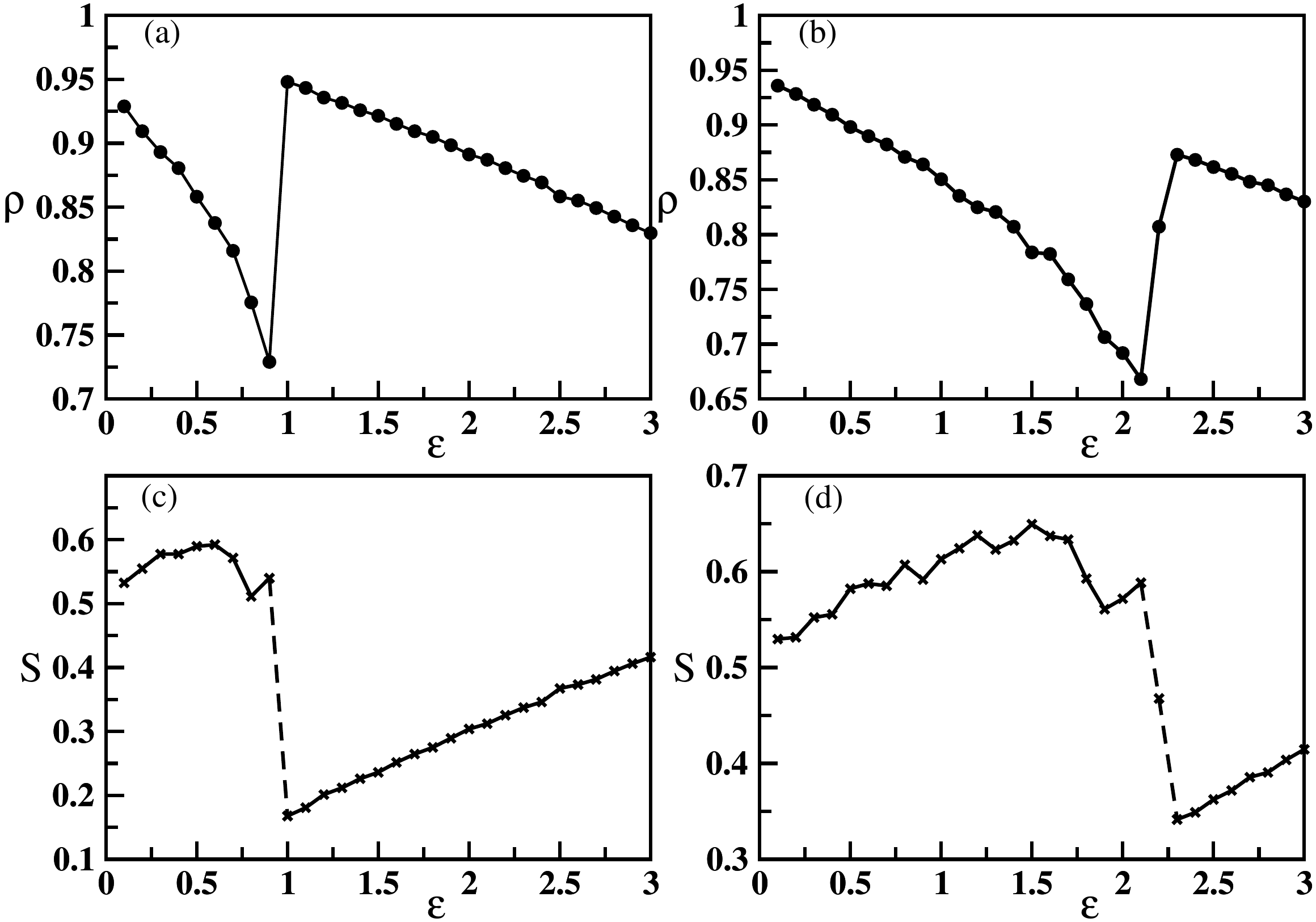}
	\end{center}
	\caption{Variation of total density ($\rho$) and {change of} entropy against $\varepsilon$ keeping all other parameters fixed. The left column [(a) \& (c)] is for $d_a=0.02$, $d_b=d_c=0.1$ and the right column [(b) \& (d)] is for $d_a=0.02$, $d_b=d_c=0.05$. The predation and reproduction rates are fixed at $p_a=p_b=p_c=0.2$, $r_a=r_b=r_c=0.4$ in both cases. (a) and (b) show the plots of $\rho$ against $\varepsilon$. (c) and (d) show the {change of} entropies of the respective final configurations. {Here $S$ is the entropy and the sudden change in $S$ has been marked in dashed lines.} {The population density shows an abrupt change at corresponding $\varepsilon_t$ indicating first-order phase transition at that point. The {change} of entropy also supports this behavior by showing a sudden change at those points.}}
	\label{single}
\end{figure*}

 {Figure}~\ref{density-single} shows the temporal behaviors of the densities of each species for different values of $\varepsilon$ keeping $d_a=0.02$ and $d_b=d_c=0.1$. In this case, for $\varepsilon<1$, we get coexistence in the long-time behavior of the density. However, the existence of single species $A$ prevails for $\varepsilon\geq 1$. So the system changes its position from coexistence to single species survival at a transition point $\varepsilon_t=1$. This value of $\varepsilon_t$ where the system changes its state is obviously dependent on other parameters (see figure~\ref{single}).

\par Now if we plot the total density, $\rho$ evaluated long time after the equilibration against the parameter $\varepsilon$, we obtain what is shown in figure~\ref{single}(a). The total density shows an abrupt change at $\varepsilon_t=1$. In figure~\ref{single}(b), we plot the same for a different set of death values and show that the abrupt change in the {density} occurs at a different value of $\varepsilon_t$.

\par {This transition occurs because it is observed in Monte-Carlo simulations that the system equilibrates towards a coexistent state although it is an unstable fixed point in view of the stability analysis \citep{bhattacharyya2020}.} For $\varepsilon<\varepsilon_t$ the system possesses two valid sets of fixed points {(i.e. possible sets of fixed points available to the system for the given parameter values)}: type II and type III. Here the system equilibrates to the type III fixed point which is the coexistent state. The fixed point densities decays with increasing $\varepsilon$ and at one stage ($\varepsilon\geq\varepsilon_t$) at least one (or more) of them becomes zero. This situation pushes the system towards the type II fixed point giving single species survival state. The switch over from one fixed point to another is captured by the abrupt change in total population density.

\par It has been observed that for further increase of $\varepsilon$ decreases the total density which is obvious. Therefore the system ultimately goes to the void state (having zero population). The transition from the single species survival state to the void state shows a continuous change in the total density. We have elaborated this part in section~\ref{phase}. 

\subsection{Transition from coexistence to void state}
\label{2nd-order}

\begin{figure*}[ht]
	\includegraphics[scale=0.3]{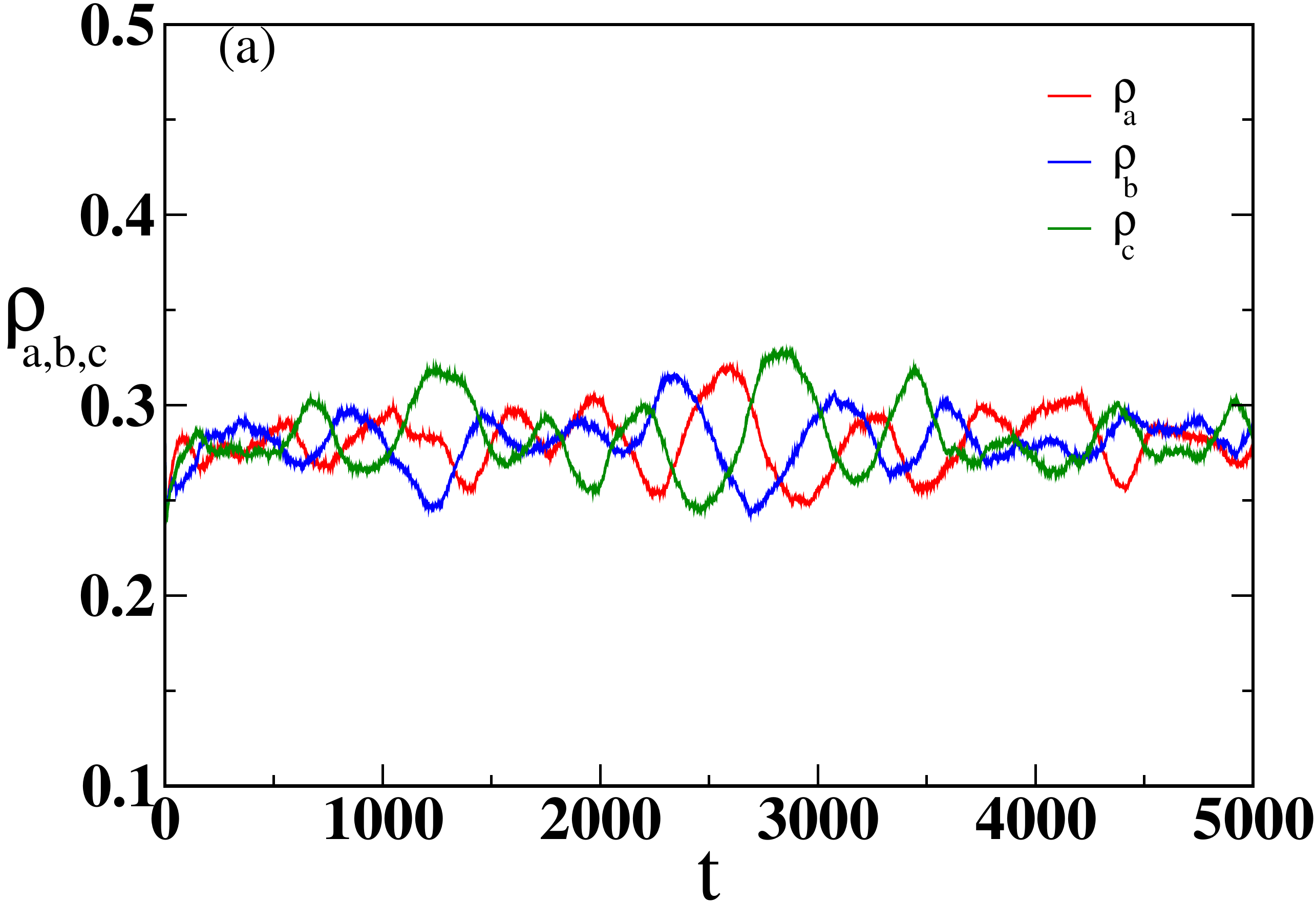}
	\includegraphics[scale=0.3]{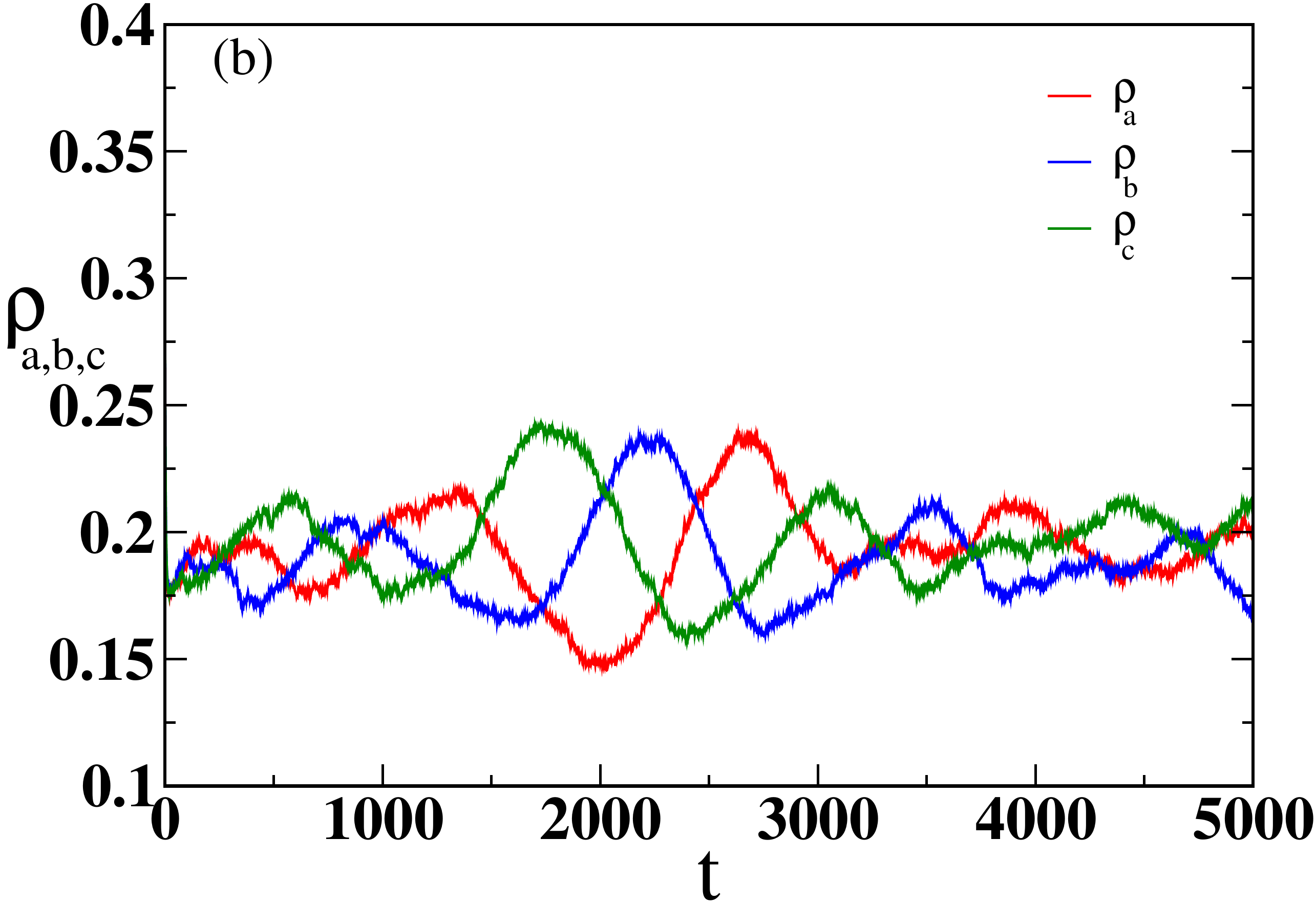}
	\caption{Density vs time plot for two different values of $\varepsilon=0.2$,0.6. Values of other parameters are fixed: $p_a=p_b=p_c=0.2$; $r_a=r_b=r_c=0.4$; $d_a=d_b=d_c=0.1$. (a) is for $\varepsilon=0.2$ and (b) is for $\varepsilon=0.6$. All the densities become zero for $\varepsilon\geq 1$ in this case.}
	\label{density-coex}
\end{figure*}

In this case, along with the previous predation and reproduction rates, we set the three death values equal as well. It ensures the coexistence of the three species initially for $\varepsilon=1$ \citep{bhattacharyya2020}. Here, {instead} of jumping to a single species survival state the system always remains in coexistence. However the total density decays with increasing $\varepsilon$. figure~\ref{density-coex} shows the time variation of the densities of the three species for different values of $\varepsilon$. It is evident that the total population decreases with increasing $\varepsilon$.

\begin{figure*}[ht]
	\begin{center}
	\includegraphics[scale=0.6]{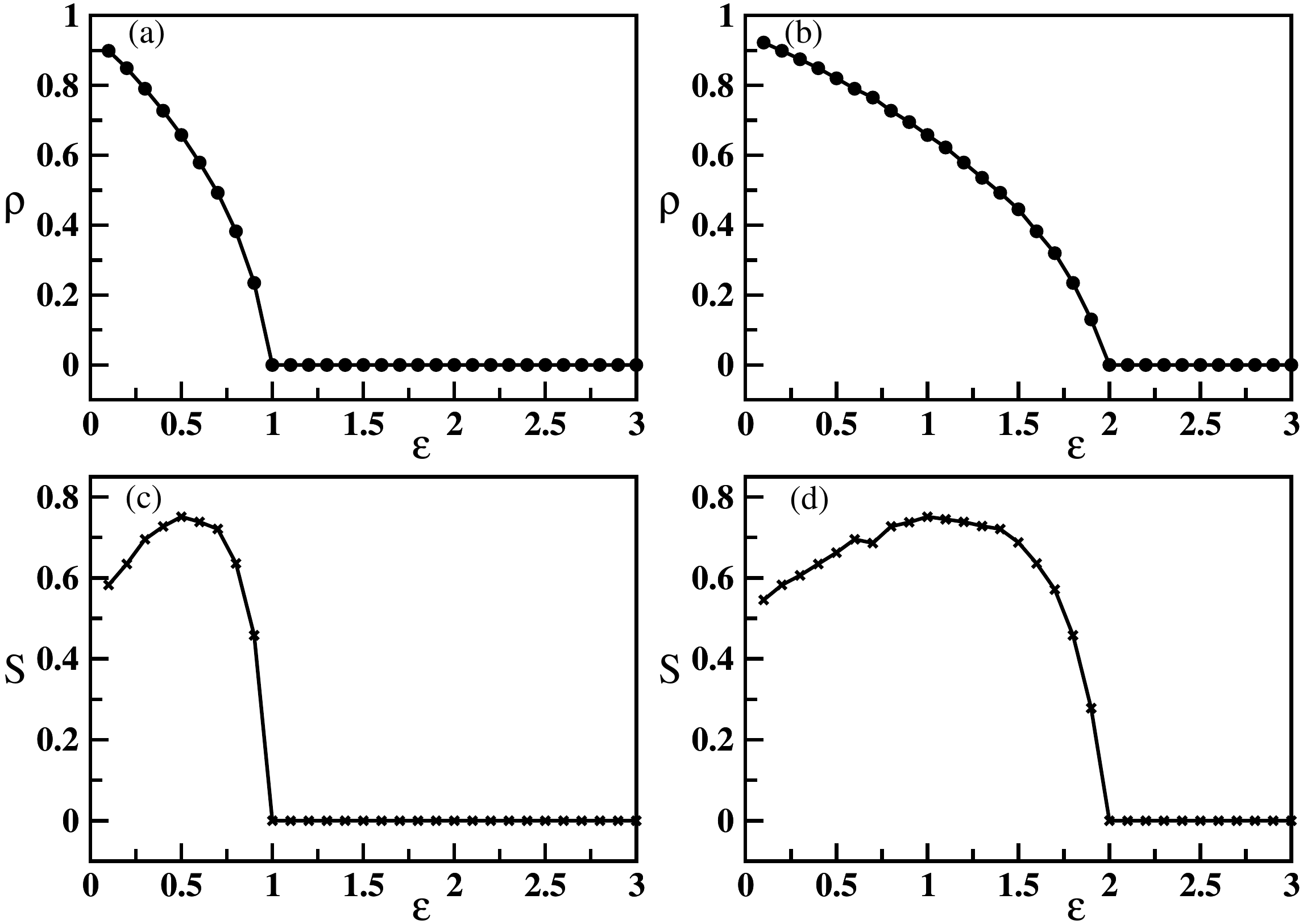}
	\end{center}
	\caption{Variation of total population density ($\rho$) and entropy ($S$) with $\varepsilon$ for two different sets of death rates. The left column [(a) \& (c)] has $d_a=d_b=d_c=0.2$ and the right column [(b) \& (d)] has $d_a=d_b=d_c=0.1$. Predation and reproduction rates are fixed to previous values. (a) and (b) show the plots of $\rho$ against $\varepsilon$. (c) and (d) show the entropies of the respective configurations at different $\varepsilon$. The density and the entropy exhibit continuous transitions at corresponding critical points, $\varepsilon_c$.}
	\label{coex}
\end{figure*}

In figure~\ref{coex} we plot the sum of the final densities for each $\varepsilon$ and find that after a certain value of $\varepsilon_c$, the total density becomes zero indicating a null state. This critical value (beyond which the population vanishes) depends on the values of other parameters. Figures~\ref{coex}(a) and \ref{coex}(b) are for different sets of such values. The variation of $\rho$ with $\varepsilon$ resembles the curve for an order parameter in 2nd order (continuous) phase transition \citep{jaeger1998}.

\par Here the system, residing in the coexistent state for $\varepsilon<\varepsilon_c$, {gradually} sheds off the fixed point densities with increasing $\varepsilon$. All these densities become zero for $\varepsilon\geq\varepsilon_c$ denoting the null state which is again the trivial fixed point of type I.

\par The behavior of the population density of the model ecosystem resembles both the 1st order and 2nd order phase transitions in two different situations. In the field of thermodynamics and statistical physics, Ehrenfest's classification of the phase transitions was based on the behavior of the thermodynamic free energy \citep{jaeger1998}. According to this the phase transitions were categorized by the lowest discontinuous derivative of the free energy at the transition point. 1st order phase transitions exhibit a discontinuity in the first derivative of the free energy, one of which is the entropy of the system. In case of the freezing of water, entropy changes discontinuously at the transition point. {On the other hand, second-order phase transitions show continuous change in entropy across the critical point.} However, the specific heat i.e. the {second derivative} of free energy has discontinuity at that point. In all these cases a measurable quantity is marked as order parameter which differentiates two phases by its value. The density difference in case of freezing of water or the magnetization in case of magnetic material are the order parameters of respective transitions. In similar manner the population density in our present work may be designated as the order parameter of both the transitions of state.\\

{
\subsection{Basin Entropy}}
\label{basin-entropy}
\noindent
{The notion that the jump from coexistence to single species in case of first-order transition and from coexistence to void state in case of second-order transition can somehow be captured by the symmetry of the distribution of species motivates to connect it with the change of basin entropy. To look into the similarity explicitly, let us calculate the basin entropy of our model ecosystem.} Following previous works \citep{daza2016,mugnaine2019}, we can calculate the basin entropy in case of our model ecosystem. For this we divide the entire system into $L$ number of blocks and calculate the entropy in each block by
\be S_l = -\sum\limits_{i=1}^{4} p_i\;\log p_i \ee
Here the sum runs over the three species and the vacancy and $p_i$ is the probability of occupation of each species or vacancy within one block. The basin entropy is obtained by the average entropy for all blocks within the system, i.e.
\be S = \dfrac{1}{L} \sum\limits_{l=1}^{L} S_l\ee
Although in principle the basin entropy should be dependent on the size of the blocks, there is no qualitative difference among them. Therefore we show the entropy obtained by dividing the systems into {$25\times25$ blocks}. The results are shown in both cases (sections~\ref{1st-order} and \ref{2nd-order}) in figures~\ref{single}(c),(d) and figures~\ref{coex}(c),(d) respectively. 
In this system, the entropy is maximum when all the species including vacancy has equal occupation probability. On the other hand the entropy becomes zero when all the sites are occupied by one species (or vacancy) only. In terms of symmetry, the former state is less symmetric than the later one. In figures~\ref{single}(c) and (d) we observe an abrupt decrement in entropy {(marked in dashed lines)} near $\varepsilon=\varepsilon_t$ as the system moves to single species existence. {The sharpness of this sudden change in the entropy can be observed more precisely in figures~\ref{app-single}(c) and (d) in Appendix~A}. Therefore this can be taken as the discontinuous change as observed in the entropies of all first-order phase transitions. One may note that the entropy does not become zero as in the single species state there are {vacant} sites also. However, in figures~\ref{coex}(c) and (d) the entropy shows continuous transition towards zero at $\varepsilon=\varepsilon_c$. {That the change is continuous can be more clearly understood from figures~\ref{app-coex}(c) and (d) in Appendix~A.} The above behavior of entropy, as defined in this system, bears its direct connection with the phenomena of 1st and 2nd order phase transitions.\\

\subsection{Phase diagram}
\label{phase}

We further explore the phase diagram for the system in the $\varepsilon-d_a$ plane. For this we keep $d_b$ and $d_c$ fixed. We observe that, if we increase $d_a$ beyond the point $d_a=d_b=d_c$, where only the continuous transitions occur, the system transits from coexistence to survival of single species $B$ and then to the void state with increasing $\varepsilon$. The first transition is discontinuous and the second transition is continuous.

\begin{figure*}[t!]
	\begin{center}
	\includegraphics[scale=0.30]{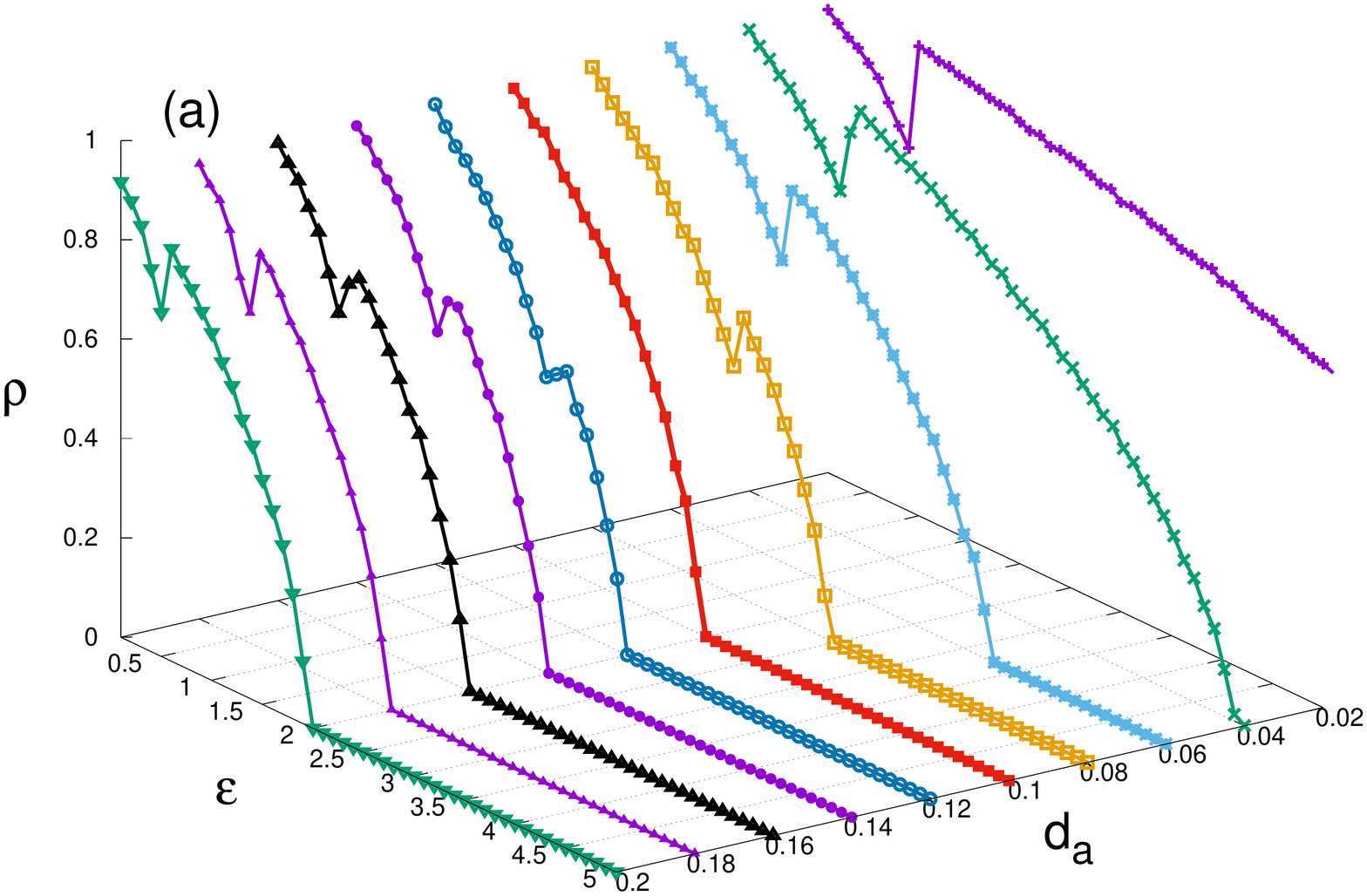}
	\includegraphics[scale=0.30]{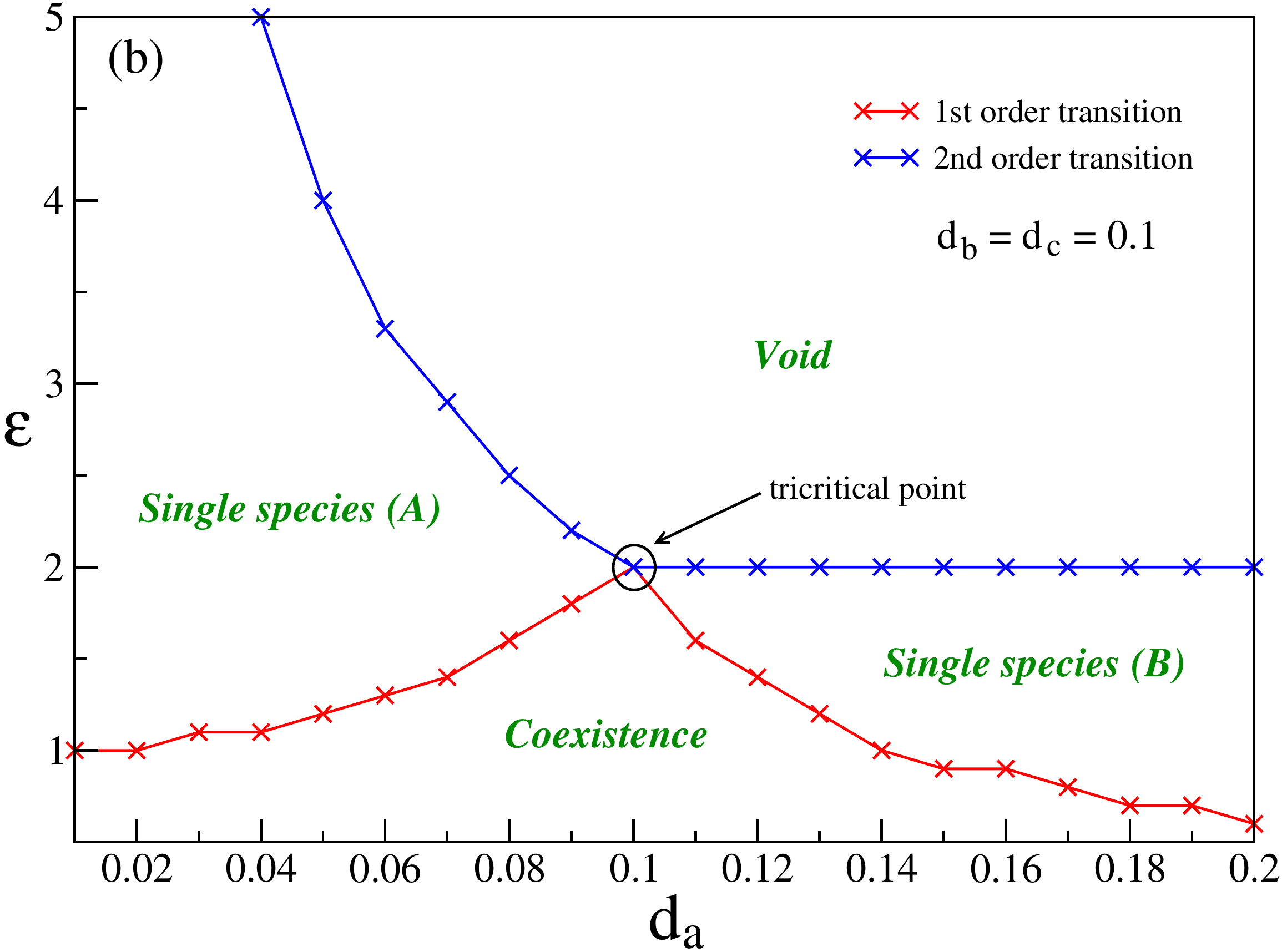}
	\end{center}
	\caption{(a) $3$D plot of $d_a-\varepsilon-\rho$ for fixed $d_b$ and $d_c$ ($d_b=d_c=0.1$). The sudden jumps observed are first-order transitions. The second-order transitions occur when the densities become zero. The red colored line for $d_a=0.1$ has only second-order transition at $\varepsilon=2.0$. (b) Phase diagram of the system in the plane of $d_a$ and $\varepsilon$ for $d_b=d_c=0.1$. The line in red color shows the first-order transition and the blue colored line shows the second-order transition. The two lines touch at the tricritical point ($d_a=0.1$, $\varepsilon=2.0$). The four phases are written in green color. These diagrams have been obtained from numerical simulations on $50 \times 50$ lattices and it has been further checked that the results do not change for higher system sizes.}
	\label{diagram3D}
\end{figure*}
\par Therefore when $d_a<d_b=d_c$, the system moves from coexistence to existence of single species $A$ via first-order transition and then from single species $A$ to void state via second-order phase transition with increasing $\varepsilon$. Similar phenomena occurs for $d_a>d_b=d_c$ with only the phase of single species $A$ being replaced by that of $B$. Right at the point $d_a=d_b=d_c$, the system has only second-order phase transition from coexistence to void state. The complete picture can be perceptible from the $3$D plot in figure~\ref{diagram3D}(a) showing variation of $\rho$ with $\varepsilon$ and $d_a$ for a fixed value of $d_b$ and $d_c$ ($d_a=d_b=0.1$). Here the discontinuous jumps in $\rho$ indicate the first-order transitions and the second-order transitions are marked by $\rho$ becoming zero. A phase diagram on the $d_a-\varepsilon$ plane corresponding to the $3$D plot is also drawn in figure~\ref{diagram3D}(b). This phase diagram shows that the point $d_a=d_b=d_c$ ($=0.1$ here) is a tricritical point where the first-order and the second-order lines meet having coexistence of three phases (coexistence, single species survival and void state).\\

\section{Discussion and outlook}
\label{conclu}
\noindent
The aim of our present study was to explore the role of a global impact that affects all constituent species in an ecosystem in a similar way. We therefore assigned a single parameter which can control all the death rates in a three-species ecosystem modeled by cyclically dominating rock-paper-scissor interaction. {Considering the total population density as the main observable, we find this quantity showing two different types of phase transitions (first-order and second-order) with the change of the controlling parameter in two different regimes of death rates.} The first-order phase transition drives the system from coexistence to single species survival state beyond a transition point whereas the system goes to a zero-population state beyond a critical value through the second-order phase transition. Based on these observations, the environmental hazards appears to have twofold impact on the ecosystem: Besides {diminishing} densities of the species the coexistence may be suddenly jeopardized or the entire population may be abolished gradually.
\par {Although the deterministic equations sometimes yield more than one set of viable fixed point for a particular parameter set, it is the stochasticity of the system that ultimately decides which fixed point to go for. Thus the stochasticity also plays the main role in the transition from one fixed point to another, similar to the role of fluctuation in triggering phase transition. It is also intriguing} to find an underlying similarity between two {apparently} different mechanisms: population in an ecosystem and phase transition in a thermodynamic system. {Symmetry within the distribution of the species plays the connecting role through the entropy in both the systems. It is percievable that when the system transits from a coexistent state to single species state the change is abrupt because of two of the species being completely eradicated and the remaining one having considerable density. On the other hand, when the system transits directly to a void state from coexistence, densities of all the species decrease gradually to zero. This change seems to be continuous therefore. The quantity that can somehow relate to both these phenomena must have some connection with how all the species are distributed within the system. That is why the basin entropy which involves the occupation probabilities of the species, shows behavioral similarity in these circumstances.} In case of an ecosystem the occupation probabilities of all the species being comparable enhances the entropy and hence affects the symmetry of the system. It is enhanced once the system faces dominance of one species or the vacancy. Therefore, in an ecosystem also, we can think of one kind of symmetry breaking to culminate in the phase transition. Thus the understanding of collective behavior in matter at thermal equilibrium, especially the concept of phase transition, surprisingly provides our cognition on some behavioral changes observed in Earth's ecosystem as well. The study made so far also provides room for further investigation in future on the behavior of species correlations in different phases. Search for any kind of universality class can also be made for further characterization of the system.\\

{
\section*{Appendix A}
\renewcommand{\theequation}{A\arabic{equation}}
\renewcommand{\thefigure}{A\arabic{figure}}
\setcounter{equation}{0}
\setcounter{figure}{0}
\label{appA}

\noindent
We further investigate for more precise location of the critical points $\varepsilon_t$ and $\varepsilon_c$ by giving smaller increment of $\varepsilon$ ($\delta\varepsilon = 0.01$) around the critical points. Due to the heavy numerical cost the areas close to the critical points have been probed.  The results - $\rho$ vs $\varepsilon$ graphs, basin entropy vs $\varepsilon$, both for first-order and second-order phase transitions are shown in figure~\ref{app-single} and figure~\ref{app-coex} respectively. The sharp changes both in density and basin entropy are clearly visible in case of the first-order transition whereas the second-order transition shows continuous changes in these two quantities.

\begin{figure}[h!]
	\begin{center}
	\includegraphics[scale=0.35]{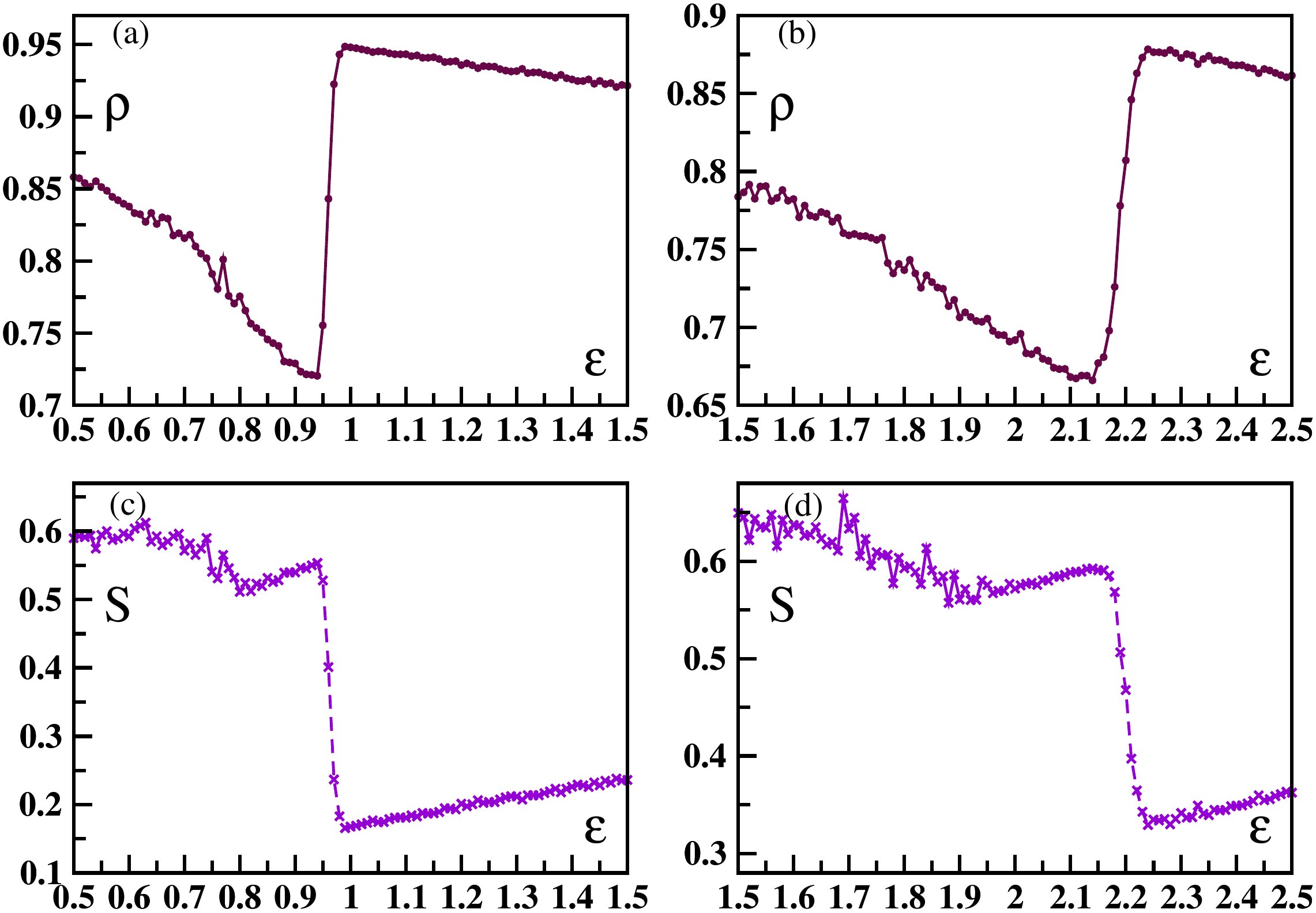}
	\end{center}
	\caption{(a) \& (b): $\rho$ vs $\varepsilon$ graph and (c) \& (d): $S$ vs $\varepsilon$ graphs close to the corresponding critical points $\varepsilon_t$. The increment in $\varepsilon$ along $x$-axis is $\delta\varepsilon=0.01$ here. The left column [(a) \& (c)] is for $d_a=0.02$, $d_b=d_c=0.1$ and the right column [(b) \& (d)] is for $d_a=0.02$, $d_b=d_c=0.05$. The predation and reproduction rates are fixed at $p_a=p_b=p_c=0.2$, $r_a=r_b=r_c=0.4$ in both cases. The abrupt changes in the entropy in (c) and (d) have been marked by dashed lines.}
	\label{app-single}
\end{figure}

\begin{figure}[h!]
	\begin{center}
	\includegraphics[scale=0.35]{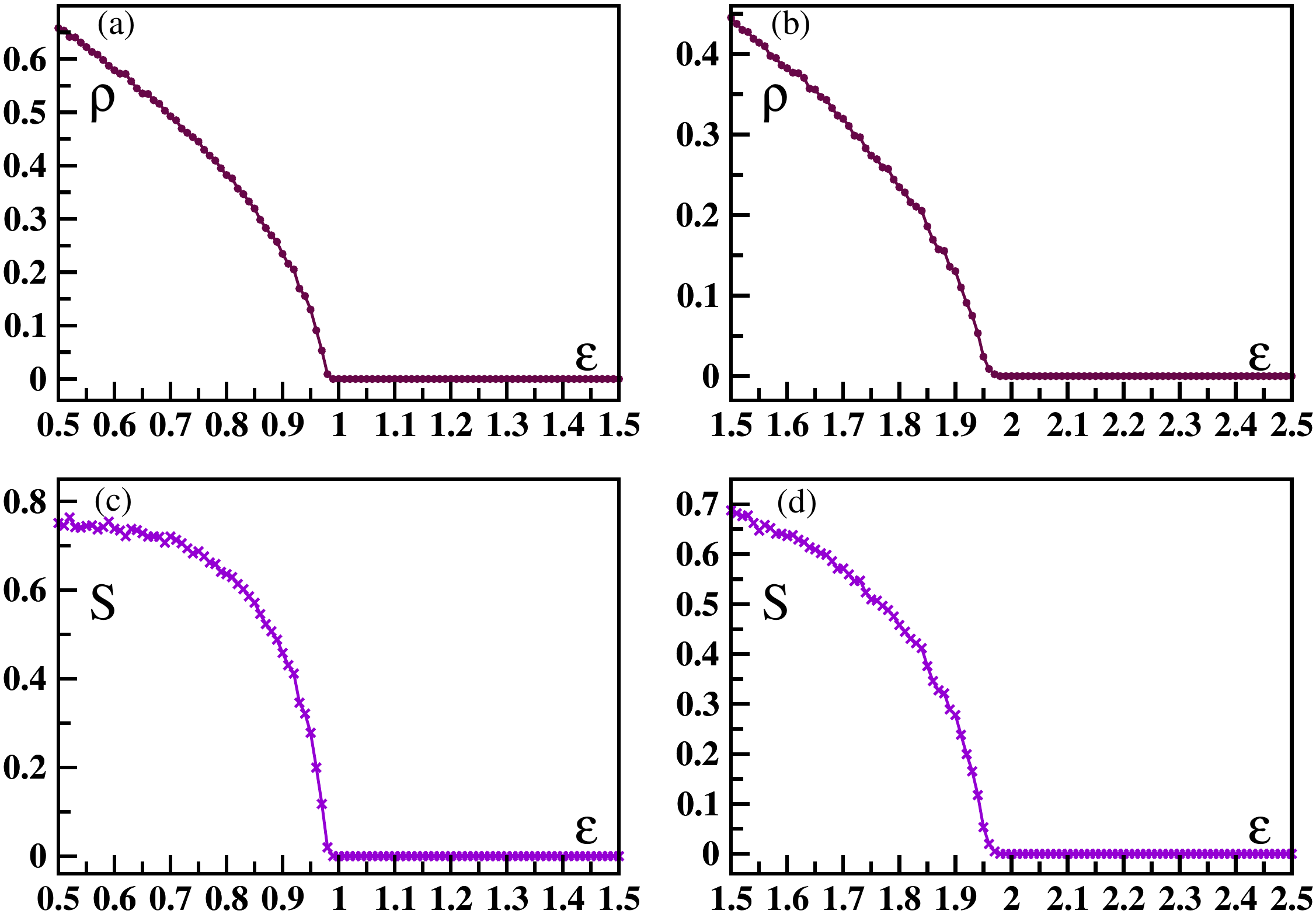}
	\end{center}
	\caption{(a) \& (b): $\rho$ vs $\varepsilon$ graph and (c) \& (d): $S$ vs $\varepsilon$ graphs close to the corresponding critical points $\varepsilon_c$. The increment in $\varepsilon$ along $x$-axis is $\delta\varepsilon=0.01$. The left column [(a) \& (c)] has $d_a=d_b=d_c=0.2$ and the right column [(b) \& (d)] has $d_a=d_b=d_c=0.1$. Predation and reproduction rates are fixed to values as in the previous figure. Even for such small increment in $\varepsilon$, the densities and the entropies exhibit continuous transitions (to zero) at critical points.}
	\label{app-coex}
\end{figure}

\section*{Acknowledgements}
\noindent
The author acknowledges anonymous referees for useful suggestions on an early version of this work.\\
}

\section*{Data availability}
\noindent
The datasets used and/or analysed during the current study is be available at \url{https://github.com/bhsirshendu/RPS-Environmental-Impact_data.git}

\section*{Conflict of interest statement}
\noindent
The author declares that he has no known competing financial interests or personal relationships
that could have appeared to influence the work reported in this paper.

\section*{ORCID iD}
\noindent
Sirshendu Bhattacharyya \url{https://orcid.org/0000-0002-7773-530X}

\bibliography{death_lib}
\bibliographystyle{apsrev4-1}

\end{document}